\documentstyle[amssymb,aps,prc]{revtex}

\begin{document}
\draft
\title{Dispersive spherical optical model of neutron scattering from $^{27}$Al up
to 250 MeV}
\author{A. Molina\thanks{%
e-mail: alberto@nucle.us.es}$^1$, R. Capote\thanks{%
e-mail: rcapotenoy@yahoo.com}$^{1,2}$, J. M. Quesada\thanks{%
e-mail: quesada@cica.es}$^1$and M. Lozano\thanks{%
e-mail: lozano@cica.es}$^1$}
\address{$^1$Departamento de F\'{\i}sica At\'{o}mica, Molecular y Nuclear,\\
Universidad de Sevilla, Facultad de F\'{\i }sica, AP 1065, E-41080 Sevilla,
Spain\\
$^2$ Centro de Estudios Aplicados al Desarrollo Nuclear, AP 100, Miramar, La
Habana, Cuba}
\date{\today}
\maketitle

\begin{abstract}
A spherical optical model potential (OMP) containing a dispersive term is
used to fit the available experimental database of $\sigma (\theta )$ and ${%
\sigma }_T$ for n+$^{27}$Al covering the energy range 0.1- 250 MeV using
relativistic kinematics and a relativistic extension of the Schroedinger
equation. A dispersive OMP with parameters that show a smooth energy
dependence and energy independent geometry are determined from fits to the
entire data set. A very good overall agreement between experimental data and
predictions is achieved up to 150 MeV. Inclusion of nonlocality effects in
the absorptive volume potential allows to achieve an excellent agreement up
to 250 MeV.
\end{abstract}


\section{Introduction}

During the last fifteen years, a great deal of theoretical attention has
been devoted to casting a proper formulation of the nuclear mean field at
positive and negative energies. A significant contribution to the solution
of this problem can be considered the work of Mahaux and co-workers on
dispersive optical-model analysis\cite
{mang84,masa87np,johoma87,masa91,masa91rev}. The unified description of
nuclear mean field in dispersive optical-model is accomplished by using a
dispersion relation, which links the real and absorptive terms of the
optical model potential. The dispersive optical model (DOM) provides a
natural extension of the optical model derived data into the bound state
region. In this way a physically self-consistent description of the energy
dependence of the OMP is obtained and the prediction of single-particle,
bound state quantities using the same potential at negative energies became
possible. Moreover additional constraint imposed by dispersion relations
helps to reduce the ambiguities in deriving phenomenological OMP parameters
from the experimental data.

Dispersive OMP analysis has been applied to nucleus-nucleus systems\cite
{namasa85,mangsa86,lo87,wafost92}, where the energy dependence of the real
central potential at low energies near the Coulomb barrier has been studied,
and contributions of the dispersion terms are evaluated. However for a
nucleus-nucleus system, the dispersive OMP analysis is limited to the
positive energy region, because it is not yet clear how to deal with
particle clusters bound in a nucleus. Some progress have been achieved in
applications of the dispersive OMP analysis to the alpha-nucleus scattering,
improving our knowledge of the alpha cluster effective interaction inside
nuclear system\cite{dalilo96}. On the other hand a great success has been
achieved in deriving DOM potentials for nucleon scattering on closed shell
nuclei like $^{40}$Ca\cite{masa91,masa88,masa89,joma88,tochde90}, $^{90}$Zr%
\cite{tochde90,dewara89,chgusm92,wafopo93,masa94} and $^{208}$Pb\cite
{mang84,masa87np,johoma87,masa91,masa88,masa89,fiwida89,wetoho96}, for which
experimental information on bound-states is available. Many studies have
dealed also with neutron scattering on non magic nuclei($^{39}$K\cite
{zhchch00},$^{51}$V\cite{lagusm89},$^{86}$Kr\cite{jocawi89},$^{89}$Y\cite
{masa87pr},$^{93}$Nb\cite{smgula86},$^{113}$In\cite{chgula90} and $^{209}$Bi%
\cite{wetoho96,dafi90,wewa99}). However very few studies have been devoted
to DOM potentials for nuclei with $A\lesssim 30$. Only one preliminary DOM
analysis has been reported for $^{27}Al(n,n)$\cite{nadeho90}. There are two
publications making a DOM analysis for proton induced reactions on aluminium
up to 60 MeV\cite{rohubo89,rosp95}.


The main purpose of this contribution is to construct a complex mean field
felt by neutrons in $^{27}$Al theoretically valid from -50 up to 250 MeV
energy. There exist two main versions of the dispersion relation approach.
In both methods, the real and imaginary parts of the mean field are
connected by a dispersion relation and, moreover, the mean field is required
to closely reproduce the experimental value of the Fermi energy $E_F$. The
main difference between the two methods is the following: (i) In the
''variational moment approach''\cite{masa88,masa89}, the parameters of the
complex mean field are determined by fitting radial moments of
phenomenological optical-model potentials.(ii) In the ''dispersive optical
model analysis''\cite{joma88,tochde90,dewara89}, the unknown parameters are
derived by performing optical-model fits to experimental scattering cross
sections that need to be available over an energy range as broad as possible.

In the present work a variation of the dispersive optical model analysis is
applied to the determination of the nuclear mean field for the neutron-$%
^{27} $Al system. An Ohio University - Los Alamos collaboration has
published an extensive survey of neutron-nucleus total cross section
measurements up to 600 MeV\cite{fifiab91,fiabfi93}. These high precision
data together with earlier neutron differential scattering data available in
the interval 1-26 MeV form the database considered at positive energies.
Fermi energy value derived from nuclear masses is used to constrain the mean
field value at negative energies. Therefore the energy variation of the
model parameters is reasonably defined over a wide range, an extremely
important point for a successful dispersive analysis. Remarkable is the fact
that our total cross section database goes up to the region where surface
absorption can be safely neglected. Since the employed database extends up
to 250 MeV and since the recent ${\sigma }_T$ data are very accurate, i.e.
the uncertainty $\Delta {\sigma }_T$ is about $\pm 1\%,$ we use relativistic
kinematics and a relativistic equivalent to the Schroedinger equation in all
our calculations.

Other motivation for our work is that aluminium is an important structural
material for the accelerator-driven systems and its cross sections are often
used as references to determine other cross sections\cite{kodebe98}. There
exist phenomenological OMP (in the sense that dispersive relations constrain
is not used) describing neutron scattering on aluminium up to high incident
energy. The LANL high energy evaluation of Chadwick {\it et al}\cite{chyo99}
employed the OMP of Petler {\it et al}\cite{peisfi85} up to 60 MeV and the
Madland global OMP\cite{ma88} from 60 up to 150 MeV. Lee and coworkers\cite
{lechfu99} derived a new phenomenological OMP which described neutron
scattering from $^{27}$Al up to 250 MeV incident energy. Recently a new
global phenomenological parametrization valid from 1 keV to 200 MeV for $%
A\geq 27$ nuclei was proposed by Koning and Delaroche\cite{kode01}.

Usually in DOM analysis the absorptive potentials are considered symmetric
about the Fermi energy $E_F$ and non-zero in the energy gap surrounding $E_F$%
. However in the contribution from Mahaux and Sartor\cite{masa91} they
pointed out that (i) due to nonlocality effects, the absorptive potential
will be highly asymmetric(with respect to $E_F$) and (ii) there should be an
energy gap centered about $E_F$ in which the absorption term drops to zero,
at least for energies between the first-hole and first-particle state.
Recent DOM analysis of neutron scattering on $^{208}$Pb and $^{209}$Bi \cite
{wetoho96} failed to describe ${\sigma }_T$ data for energies above 40 MeV
using asymmetric version of the absorptive potentials for large positive and
large negative energies. We will present strong evidence to favour
asymmetric absorptive potentials for proper description of the neutron
scattering ${\sigma }_T$ data for energies between 150 and 250 MeV.

The paper is structured as follows. Section II provides a description of the
dispersive optical model formalism, the solved wave equation and the forms
of the energy and radial dependencies of the real, imaginary and spin-orbit
potentials. Section III describes the compound nucleus (CN) calculations,
the $^{27}$Al(n,n) experimental database, our procedure for searching, and
the resulting relativistic and non-relativistic spherical DOM potentials for
$^{27}$Al(n,n). In the same section we compare derived DOM potentials with
phenomenological potentials and experimental data. Finally Section IV
contains our conclusions.

\section{DOM formalism}

\subsection{Optical-model potential and wave equation}

The optical-model analysis was carried out with a semirelativistic
generalization of the conventional nonrelativistic Schroedinger formulation
of the scattering process\cite{nascsi81}. Relativistic kinematics was used
for the projectile, but it was assumed that the target motion in the
center-of-mass system could be treated nonrelativistically. A relativistic
equivalent to the Schroedinger equation was generated by appropriate
reduction of the Dirac equation for a massive, energetic fermion (mass $m$
and c.m. wave number $k)$ moving in a localized central potential $V(r)$
taken as the time-like component of a Lorentz four-vector. In the reduced
two-body problem with relativistic projectile but nonrelativistic target
(mass $M$) the large component of the partial wave function $F_l(\rho )$ can
be shown to satisfy the radial equation

\begin{equation}
\left\{ \frac{d^2}{d\rho ^2}+\left[ 1-\frac{V(\rho )}{T_c}-\frac{l(l+1)}{%
\rho ^2}\right] \right\} F_l(\rho )=0  \label{Sch_eq}
\end{equation}
where $\rho =kr$, $T_c$ is the total c.m.kinetic energy, $l$ is the orbital
angular momentum, and $V(\rho )$ is the renormalized nuclear optical
potential

\begin{equation}
V(\rho )=\gamma U(r),\gamma =1+\frac{T_c}{T_c+2m}  \label{gamma}
\end{equation}

Equation (\ref{Sch_eq}) is formally identical to the radial equation for the
solution of the non-relativistic Schroedinger equation for the analogous
scattering problem with a nuclear potential renormalized by a factor $\gamma
$. This factor becomes increasingly important as the projectile kinetic
energy increases (see equation (\ref{gamma})) leading to an effective
increase of the potential depth. The spin-orbit term in $V(r)$ employed in
this analysis is a purely phenomenological one since the intrinsic SO term
in the Dirac equation is negligible small in the above limits. Equation (\ref
{Sch_eq}) was used in all calculations. In a non-relativistic case we set a
factor $\gamma $ equal to 1 and non-relativistic kinematics was employed,
otherwise relativistic kinematics and the factor $\gamma $ according to
equation (\ref{gamma}) were used.

Our analysis spans an energy range from 0.1 up to 250 MeV. Both direct and
statistical processes contribute to nucleon-nucleus elastic scattering at
these energies. According to our estimation the statistical processes are
important up to 12 MeV in aluminium. Compound nucleus calculation will be
described in the next section. The direct processes, increasingly dominant
at higher energies, can be described by the optical model. Although $^{27}$%
Al nucleus is deformed, the spherical OMP has been applied successfully\cite
{peisfi85,lechfu99,whdago84}. {\it A posteriori} analysis of the impact of
this approximation on the calculated observables will be discussed below.

The optical model potential may be written as

\begin{eqnarray}
U(r,E) &=&-\left[ V_v(E)+iW_v(r,E)\right] f_{_{WS}}(r,R_v,a_v)  \nonumber \\
&&-\left[ V_s(E)+iW_s(r,E)\right] g_{_{WS}}(r,R_s,a_s)  \nonumber \\
&&-\left( \frac \hbar {m_\pi c}\right) ^2\left[ V_{so}(E)+iW_{so}(E)\right]
\times \frac 1rg_{_{WS}}(r,R_{so},a_{so})\left( {\vec{l}}\cdot {\vec{\sigma}}%
\right)  \label{OMPfull}
\end{eqnarray}

\noindent
where the successive complex-valued terms are the volume central, surface
central and spin-orbit potentials. The volume shape $f_{_{WS}}(r,R_v,a_v)$
is a standard Woods-Saxon form factor specified by a potential radius $R_v$
and diffuseness $a_v$. The surface(spin-orbit) shape is the first derivative
of the Woods-Saxon form specified by a potential radius $R_s(R_{so})$ and
diffuseness $a_s(a_{so})$%
\begin{equation}
g_{_{WS}}(r,R_i,a_i)=-4a_i\frac d{dr}f(r,R_i,a_i)
\end{equation}
\noindent
The reduced radius parameter $r_i$ is introduced as usual by the relation $%
R_i=r_iA^{1/3}$. In our formulation of the OMP in Eq.(\ref{OMPfull}) the
real and imaginary central volume terms share the same geometry parameters $%
r_{v\text{ }}$and $a_v$ and likewise the real and imaginary central surface
terms share the same $r_s$ and $a_s$. This assumption\cite{johoma87} can be
seen as a consequence of the dispersive relations, allowing us to reduce the
number of geometrical parameters in the OMP.

For the spin-orbit potential we adopt the parameters obtained by Koning {\it %
et al}\cite{kodebe98}, namely:
\begin{eqnarray}
&&V_{so}(E)=6.0\exp (-0.005E)\text{ MeV}  \nonumber \\
&&W_{so}(E)=0.2-0.011E\text{ MeV}  \nonumber \\
&&r_{so}=1.017\text{ fm, }a_{so}=0.6\text{ fm}
\end{eqnarray}

In a dispersion relation treatment, the real central potential strength
consists of a term which varies slowly with energy, the so called
Hartree-Fock (HF) term, $V_{HF}(E)$, plus a correction term, $\triangle V(E)$%
, which is calculated using a dispersion relation. The depth of the
dispersive term of the potential $\triangle V(E)$ can be written in the
substracted form

\begin{equation}
\triangle V(E)=\frac{{\cal P}}\pi {\int_{-\infty }^\infty W(E^{\prime })}%
\left( \frac 1{E^{\prime }-E}-\frac 1{E^{\prime }-E_F}\right) dE^{\prime }
\label{integral_subs}
\end{equation}

With the assumption that $W(E)$ be symmetric respect to the Fermi energy $%
E_F $, Eq.(\ref{integral_subs}) can be expressed in a form which is stable
under numerical treatment\cite{dewara89}, namely:

\begin{equation}
\triangle V(E)=\frac 2\pi (E-E_F){\int_{E_F}^\infty }\frac{W(E^{\prime
})-W(E)}{(E^{\prime }-E_F)^2-(E-E_F)^2}dE^{\prime }  \label{integral}
\end{equation}
where $W(E)$ is the imaginary part of the OMP. The dispersive term $%
\triangle V(E)$ is divided into two terms $\triangle V_v(E)$ and $\triangle
V_s(E)$, which arise through dispersion relations (\ref{integral}) from the
volume $W_v(E)$ and surface $W_s(E)$ imaginary potentials respectively. If
imaginary potential geometry is energy dependent, then radial dependence of
the dispersive correction can not be expressed using a Wood-Saxon form
factor, i.e $\triangle V_v(r,E)\neq $$\triangle V(E)f(r,R,a)$. However, to
simplify the problem, the OMP geometry parameters used in this work are
energy independent. In this case, using the definitions of the equation (\ref
{OMPfull}), the real volume $V_v(E)$ and surface $V_s(E)$ central part of
the DOM potential are given by
\begin{eqnarray}
&&V_v(E)=V_{HF}(E)+\triangle V_v(E)  \nonumber \\
&&V_s(E)=\triangle V_s(E)
\end{eqnarray}

It is known that the energy dependence of the depth $V_{HF}(E)$ is due to
the replacement of a microscopic nonlocal HF potential by a local
equivalent. For a gaussian non-locality $V_{HF}(E)$ is a linear function of $%
E$ for large negative $E$ and is an exponential for large positive $E$.
Following Mahaux and Sartor\cite{masa91}, the energy dependence of the
Hartree-Fock part of the nuclear mean field is taken as that found by
Lipperheide\cite{li67}:
\begin{equation}
V_{HF}(E)=V_0\exp (-\alpha _{_{HF}}(E-E_F))  \label{VHF}
\end{equation}
where the parameters $V_0$ and $\alpha _{_{HF}}$ are undetermined constants.
This equation (\ref{VHF}) can be used to describe HF potential in the
scattering regime\cite{masa91}.

It is useful to represent the variation of surface $W_s(E)$ and volume
absorption potential $W_v(E)$ depth with energy in functional forms suitable
for the dispersive optical model analysis. An energy dependence for the
imaginary volume term has been suggested in studies of nuclear matter theory%
\cite{brrh81}:

\begin{equation}
W_v(E)=A_v~\frac{(E-E_F)^n}{(E-E_F)^n+(B_v)^n}  \label{volumen}
\end{equation}
where $A_v$ and $B_v$ are undetermined constants. Following Mahaux and Sartor%
\cite{masa87np} we adopt $n=4$. An energy dependence for the
imaginary-surface term has been suggested by Delaroche {\it et al}\cite
{dewara89} to be:

\begin{equation}
W_s(E)=A_s~\frac{(E-E_F)^m}{(E-E_F)^m+(B_s)^m}~\exp (-C_s|E-E_F|)
\label{surface}
\end{equation}
where $m=4$ and $A_s,B_s$ and $C_s$ are undetermined constants.

According to equations (\ref{volumen}) and (\ref{surface}) the imaginary
part of the OMP is assumed to be zero at $E=E_F$ and nonzero everywhere
else. A more realistic parametrization of $W_v(E)$ and $W_s(E)$ forces these
terms to be zero in some region around the Fermi energy. A physically
reasonable energy for defining such a region is the average energy of the
single-particle states $E_P$\cite{masa91}. For aluminium we used a value $%
E_P=-5.66$ MeV, obtained by averaging the first three particle states
reported in the microscopical single-particle level calculation by Moller
and Nix\cite{RIPL98}. The experimental value of the Fermi energy $E_F$
derived from mass differences is equal to -10.392 MeV.

Therefore a new definition for imaginary part of the OMP\ can be written as:

\[
W_v(E)=\left\{
\begin{array}{cl}
0 & {\text{ for }E_F<E<E_p} \\
A_v~{\displaystyle {\frac{(E-E_p)^n }{(E-E_p)^n+(B_v)^n}}} & {\text{ for }%
E_p<E}
\end{array}
\right.
\]
\begin{equation}  \label{WV3}
\end{equation}

\noindent and likewise for surface absorption.

\[
W_s(E)=\left\{
\begin{array}{cl}
0 & {\text{ for }E_F<E<E_p} \\
A_s~{\displaystyle {\frac{(E-E_p)^m }{(E-E_p)^m+(B_s)^m}}~\exp(-C_s|E-E_P|)}
& {\text{ for }E_p<E}
\end{array}
\right.
\]
\begin{equation}  \label{WS3}
\end{equation}

The symmetry condition
\begin{equation}
W(2E_F-E)=W(E)  \label{symmetry}
\end{equation}
is used to define imaginary part of the OMP for energies below the Fermi
energy. Equations (\ref{WV3}) and (\ref{WS3}) are used to describe imaginary
absorptive potential in this contribution.

\subsection{High energy behavior of the volume absorption}

The assumption that imaginary potential $W_v(E)$ is symmetric about $%
E^{\prime }=E_F$ (according to equation (\ref{symmetry})) is plausible for
small values of $\left| E^{\prime }-E_F\right| $, however as was pointed out
by Mahaux and Sartor\cite{masa91} this approximate symmetry no longer holds
for large values of $\left| E^{\prime }-E_F\right| $. In fact the influence
of the nonlocality of the imaginary part of the microscopic mean field will
produce an increase of the empirical imaginary part $W(r,E^{\prime })$ at
large positive $E^{\prime }$ and approaches zero at large negative $%
E^{\prime }$\cite{mang84,masa86}. Following Mahaux and Sartor\cite{masa91},
we assume that the absorption strengths are not modified below some fixed
energy $E_a$. They used $E_a=60$ MeV, however this value is fairly arbitrary%
\cite{masa91}. Let assume the non-local imaginary potential to be used in
the dispersive integral is denoted by $\widetilde{W}_v(E)$, then we can write%
\cite{masa91rev}
\begin{equation}
\widetilde{W}_v(E)=W_v(E)\left[ 1-\frac{(E_F-E-E_a)^2}{(E_F-E-E_a)^2+E_a^2}%
\right] \text{, for }E<E_F-E_a  \label{WVnonlocal1}
\end{equation}
and
\begin{equation}
\widetilde{W}_v(E)=W_v(E)+\alpha \left[ \sqrt{E}+\frac{(E_F+E_a)^{3/2}}{2E}-%
\frac 32\sqrt{(E_F+E_a)}\right] \text{, for }E>E_F+E_a  \label{WVnonlocal2}
\end{equation}
These functional forms are chosen in such a way that the function and its
first derivative are continuous at $E^{\prime }=\left| E_F-E_a\right| $. At
large positive energies nucleons sense the ''hard core'' repulsive region of
the nucleon-nucleon interaction and $\widetilde{W}_v(E)$ diverges like $%
\alpha \sqrt{E}$. Using a model of a dilute Fermi gas hard-sphere the
coefficient $\alpha $ can be estimated to be equal to 1.65 MeV$^{1/2}$\cite
{masa86}, assuming that the Fermi impulse $k_{F\text{ }}$is equal to $1.36$
fm$^{-1}$and the radius of the repulsive hard core is equal to $0.4$ fm. On
the contrary, at large negative energies the volume absorption decreases and
goes asymptotically to zero. The non-local imaginary absorption potential $%
\widetilde{W}_v(E)$ and the symmetric imaginary absorption potential $W(E)$
are represented by solid and dotted lines respectively in the lower panel of
figure \ref{FIG_WDOM}.

The asymmetric form of the volume imaginary potential of equations (\ref
{WVnonlocal1}) and (\ref{WVnonlocal2}) results in a dispersion relation that
must be calculated directly from Eq.(\ref{integral_subs}) and separates into
three additive terms\cite{vaweto00}. Therefore, we write the dispersive
correction in the form

\begin{equation}
\triangle \widetilde{V}_v(E)=\triangle V_v(E)+\triangle V_{<}(E)+\triangle
V_{>}(E)
\end{equation}
where $\triangle V_v(E)$ is the dispersive correction due to the symmetric
imaginary potential of equation (\ref{WV3}) and the terms $\triangle
V_{<}(E) $ and $\triangle V_{>}(E)$ are the dispersive corrections due to
the asymmetric terms of equations (\ref{WVnonlocal1}) and (\ref{WVnonlocal2}%
), respectively. The resulting energy dependence of the dispersive integrals
$\triangle \widetilde{V}_v(E)$ and $\triangle V_v(E)$ for both non-local
imaginary absorption potential $\widetilde{W}_v(E)$ and symmetric imaginary
absorption potential $W(E)$ is represented by solid and dotted lines
respectively in the upper panel of figure \ref{FIG_WDOM}. While the
symmetric case features an equal contribution coming from negative and
positive energies, in the asymmetric case the negative energy contribution
to the dispersive integral is very different to the positive energy value.
The resulting dispersive correction for the asymmetric case starts to
increase already for energies above 50 MeV, making a significant
contribution to the real part of the OMP.

It should be noted that non-locality corrections (equations (\ref
{WVnonlocal1}) and (\ref{WVnonlocal2})) can be used either for volume or
surface imaginary potential; however, Mahaux and Sartor\cite{masa91} have
shown that nonlocality consideration for the surface imaginary potential has
a very small effect on calculated cross sections. Therefore in this work we
followed Ref.\cite{masa91rev} and only considered the effects of nonlocality
in the volume absorption.

\section{Dispersive optical model analysis}

\subsection{DOM software}

Search optical model codes ECIS95 in the external input mode\cite
{ra72,ecis96} and COH v 2.2\cite{COH} were used for DOM analyses using
relativistic and non-relativistic kinematics respectively. A modification
was introduced into the later code to force equality of the real and
imaginary surface(volume) geometry parameters $R_s,a_s(R_v,a_v)$ during the
search procedure, as it is implicit in the equation (\ref{OMPfull}). The
code does not include the dispersion relations, therefore the dispersion
integrals (\ref{integral}) of the symmetric forms (\ref{WV3}) and (\ref{WS3}%
) of the imaginary potential were calculated numerically using a Gauss
quadrature method\cite{camoqu01}, while the asymmetric contribution was
calculated analytically (see Eqs.(16-19) of Ref.\cite{vaweto00}). An
auxiliary code system was developed to produce proper input data sets for
both optical model codes and to calculate for each data set(i.e. for each
energy) the $\chi ^{2\text{ }}$quantity according to:

\begin{equation}
\chi ^2(E)={\sum_{i=1}^{N_\sigma }}\left[ \frac{\sigma _{\exp }(E,\theta
_i)-\sigma _{calc}(E,\theta _i)}{\Delta \sigma _{\exp }(E,\theta _i)}\right]
^2+\left[ \frac{\sigma _{\exp }^{tot}(E)-\sigma _{calc}^{tot}(E)}{\Delta
\sigma _{\exp }^{tot}(E)}\right] ^2
\end{equation}

Here, $\sigma _{calc}(E,\theta _i)[\sigma _{calc}^{tot}(E)]$ and $\sigma
_{\exp }(E,\theta _i)$ [$\sigma _{\exp }^{tot}(E)$], are the
differential(total) cross sections from the optical model calculations and
experiments for a given laboratory energy $E$, respectively, and $\Delta
\sigma _{\exp }(E,\theta _i)[\Delta \sigma _{\exp }^{tot}(E)]$ is the
experimental uncertainty reported. The $N_\sigma $ is the number of data
points for $\sigma _{\exp }(E,\theta _i)$. Our code system allows to
finetune the OMP parameters of interest to minimize the total search $\chi
^{2\text{ }}$of the entire data set.

\subsection{Summary of the experimental databases}

A survey of the experimental data spanning from 0.1 to 250 MeV used in the
DOM analyses is presented in this section. The $^{27}$Al(n,n) $\sigma
(\theta )$ data were obtained from Towle and Gilboy\cite{togi62} at 1, 2,3
and 4 MeV; Tanaka {\it et al}\cite{tatsma69} at 4.8,6,7 and 8 MeV; Kinney
and Perey\cite{kipe70} at 5.4,6.4,7.5 and 8.6 MeV; Dagge {\it et al}\cite
{dagrha89} at 7.62 MeV; Velkley {\it et al}\cite{veglbr74} at 9 MeV; Boerker
{\it et al}\cite{bobobr88} at 10.2 MeV; Whisnant {\it et al}\cite{whdago84}
at 11,14 and 17 MeV; Petler {\it et al}\cite{peisfi85} at 18,20,22,25 and 26
MeV; Bratenahl {\it et al}\cite{brfehi50} at 84 MeV; Salmon\cite{sa60} at 96
MeV and Van Zyl {\it et al}\cite{vavowi56} at 136 MeV. The $^{27}$Al(n,n) $%
A_y(\theta )$ data were obtained from Dagge {\it et al}\cite{dagrha89} at
7.62 MeV and Martin and Walter\cite{mawa86} at 14 and 17 MeV. These
polarization data were used only for testing spin-orbit interaction.
Energy-averaged total cross sections ${\sigma }_T$ for $^{27}$Al were
obtained from Finlay {\it et al}\cite{fifiab91,fiabfi93} from 5.3 to 250
MeV. Additional energy-averaged ${\sigma }_{T\text{ }}$ data were taken from
Refs.\cite
{demo51,tawo53,boscst61,mepa66,scco68,cifoko68,fogl71,peloki72,lahahi81,frgrle88,oh85,roshna94}
to be used for comparing predictions of the model. We selected measurements
containing several points in energy, specially all with data above 20 MeV.
In critiquing all the available experimental total cross section data, the
high resolution cross section data of Ref.\cite{scsche74} was found to be
inconsistent with the rest of the data set and was ignored in our analysis.

\subsection{Compound-nucleus corrections}

The statistical model of nuclear reaction according to the Hauser-Feshbach
theory\cite{hafe52} with width fluctuation corrections as modified by
Moldauer\cite{Mo80} is used to compute the CN contributions to the elastic
channel. When the cross section is averaged over many CN resonances the
shape elastic differential cross section can be incoherently added to the
compound elastic contribution to compare with the experimentally observed
elastic-scattering cross section. For neutron energies larger than 12 MeV,
the compound-elastic contribution can be neglected. The CN cross section
calculation is built-in inside the search code CoH\cite{COH}. Three reaction
channels are considered in the statistical-model calculations of the $%
^{28}Al $ CN decay: (n,n), (n,p) and (n,$\alpha $). Transmission
coefficients for proton and alpha emission in the exit channels are
calculated from the spherical OMP parameters by Perey {\it et al}\cite{pe63}
and Arthur and Young\cite{aryo80}(a modification of Lemos OMP\cite{le76})
respectively. The transmission coefficients in the entrance and inelastic
channel were calculated using the DOM potential of the present work.

Discrete level information is used to represent the low-lying states and the
Gilbert-Cameron level density formulae\cite{gica65} are used to represent
the high-lying continuum of states. Figure \ref{FIG_CUMUL_LEV} shows the
cumulative number of levels as a function of excitation energy for the
residual nuclei of the three reaction channels. The discrete state data are
taken from the Belgya compilation contained in RIPL\cite{RIPL98}. The
vertical lines indicate the cut-off energy between the discrete states and
the continuum. It is well known that a CN calculation is highly sensitive to
the level density parameters modeling the continuum of the excited states.
We used the ''constant temperature'' formula \cite{gica65} to estimate the
total number of excited states available at excitation energy $E$, $%
N(E)=exp((E-E_0)/T)$, where $T$ is the ''nuclear temperature'' and $E_{0%
\text{ }}$is the energy shift. These two parameters are determined by
fitting the cumulative number of available experimental states up to some
cut-off energy. The level density parameters for all three residual nuclei
involved in CN cross section calculations are listed in Table \ref
{LEV_DENS_PARAM}. Cumulative number of levels as calculated by the
''constant temperature'' model using these parameters is shown as solid
lines in figure \ref{FIG_CUMUL_LEV}.

\subsection{Search procedure}

It is well known that search routine does not always converge on optimum
solution specially when we are dealing with strongly correlated OMP
parameters. In our DOM analysis we performed a global $\chi ^2$ optimization
combined with a grid search using a $\chi ^{2\text{ }}$fit in a limited
energy region using a maximum number of two fitting parameters
simultaneously. Our search procedure can be divided in four main steps:{\ }

\begin{enumerate}
\item  {Search for imaginary $W_v^{emp}(E)$ empirical potential depth using
total cross section data between 70 and 150 MeV, neglecting real and
imaginary surface contribution. This energy range is selected in order to
neglect surface absorptive potential in the first iteration. Once empirical
values $W_v^{emp}(E)$ were obtained a fit of the absorptive volume potential
$W_v(E)$ using equation (\ref{WV3}) is carried out. In this way volume
absorption is fixed, as well as dispersive volume contribution $\triangle
V_v(E)$ to the central real potential, which is calculated by integration.
The empirical values of the real volume potential depth $V_v^{emp}(E)$
combined with the $\triangle V_v(E)$ are used to obtain a set of empirical
points corresponding to $V_{HF}^{emp}(E)$. A typical set of empirical values
derived in the above described way can be seen in figure \ref{FIG_WVOL_FIT},
as obtained with the search code COH. Finally the equation (\ref{VHF}) is
used to obtain the $V_0$ and $\alpha _{_{HF}}$ parameters that offer a best
fit to the empirical real potential data. In the fitting process the
strength $V_0$ was constrained for the DOM predicted first-particle and
first-hole states to be centered around the experimental value of the Fermi
energy. }

\item  {\ At each energy for which neutron elastic differential cross
section and neutron total cross section data are available from 1 up to 26
MeV, we have conducted a best $\chi ^2$ fit by searching on volume real $%
V_v^{emp}(E)$ and surface imaginary $W_s^{emp}(E)$ empirical potential
depths. In the first iteration the corresponding dispersive surface
contribution $\triangle V_s(E)$ to the central real potential was calculated
by integration from the starting OMP parameters. CN contributions and width
fluctuations corrections were considered in all calculation for incident
energy below 12 MeV. Once empirical values $W_s^{emp}(E)$ were obtained a
fit of the absorptive surface-peaked potential $W_s(E)$ using equation (\ref
{WS3}) is carried out. Dispersive surface contribution $\triangle V_s(E)$ to
the central real potential is re-evaluated by integration. The empirical
values of the real volume potential depth $V_v^{emp}(E)$ combined with the $%
\triangle V_v(E)$ calculated for these energies are used to increase the set
of empirical points corresponding to $V_{HF}^{emp}(E)$. The equation (\ref
{VHF}) is used to refine the fitting of the $V_{0\text{ }}$and $\alpha _{HF}$
parameters, derived in point (1), using the whole empirical set of potential
values obtained in steps (1) and (2). We iterate over steps (1) and (2)
until the empirical potential strengths were consistent with our predefined
energy functional (see equations (\ref{VHF}),(\ref{WS3}) and (\ref{WV3}))
over the whole energy range }

\item  {\ After fixing potential strengths, the optimum geometry parameters
were searched for, iterating over steps (1) and (2) to redefine the
potential strengths corresponding to the optimized geometry parameters }

\item  {\ Finally a global $\chi ^2$optimization using the whole
experimental database was carried out to obtain the minimum in the $\chi ^2$
multiparameter surface.}
\end{enumerate}

\subsection{\protect\smallskip \protect\smallskip The $^{27}Al(n,n)$\ DOM
analysis}

We started our analysis by using non-relativistic formulation to fit the
experimental data. Initial values for geometrical parameters were provided
by the energy independent geometry deduced by Whisnant {\it et al}\cite
{whdago84} and used{\it \ }by{\it \ }Petler {\it et al}\cite{peisfi85} for
phenomenological analysis of the data up to 26 MeV. They found $r_v=1.18$
fm, $a_v=0.64$ fm and $r_s=1.26$ fm, $a_s=0.58$ fm$.$ Because the general
form of the energy dependence of the imaginary potential used in the present
model is similar to the $^{27}Al(n,n)$ phenomenological OMP of Lee {\it et al%
}\cite{lechfu99}, we used their volume real and imaginary potential
parameters as a starting point for our analysis. We were using symmetric
imaginary absorptive potentials according to equations (\ref{WS3}) and (\ref
{WV3}), therefore we adjusted 7 parameters, namely $(V_{0\text{ }},\alpha
_{_{HF}})$, which define the smooth energy dependence of the real volume
potential and $(A_v,B_v$ $)$ and $(A_s,B_s,C_s)$ defining the volume and
surface absorptive potential respectively. After proper values were obtained
by this global minimization the energy independent geometry parameters were
also optimized. The derived non-relativistic DOM potential parameters are
listed in Table \ref{OMP_NOREL_PARAM}.

The final $\sigma _T$ DOM fits using non-relativistic potential are compared
to $^{27}Al(n,n)$ data in Fig.\ref{FIG_SIGMA_HIGH_E_NONREL}. It should be
stressed that experimental total cross section data(except the grayed one)
shown in this figure were not used in the DOM parameter search. We can
observe that the experimental total cross section at energies above 130 MeV
was always underestimated by our nonrelativistic calculations. We can not
change the real volume potential depth (or the so called Hartree-Fock
potential) without spoiling the fits to the differential cross section. One
solution could be to consider an increase of the radius of real part of the
OMP. However this approach would obscure our treatment with energy
independent geometry. Furthermore, it is theoretically obvious, that
relativistic effects and non-locality should show up at this energy regime.
Therefore we decided to carry out a full relativistic treatment, including
non-local contribution to the absorptive potential, which will be reflected
on the dispersive contribution to the real potential.

The starting point in this second stage was the non-relativistic DOM
potential. We took into account the non-local contribution to the volume
absorptive potential according to equations (\ref{WVnonlocal1}) and (\ref
{WVnonlocal2}). Only one additional parameter was included, namely the
energy $E_a$ above which the non-local behavior of the volume absorptive
potential is considered. In this later {$\chi ^2$ }minimization the total
cross section data up to 250 MeV were included into the experimental
database. All potential parameters changed because of the sizeable
contribution of the non-local absorption for energies above 40\ MeV as can
be seen from figure \ref{FIG_REL_NONLOCAL_CONTRIB}. In the same figure the
total cross section calculated with the non-relativistic DOM potential is
shown for comparison. It is interesting to remark that relativistic
correction alone is clearly not enough for the correct description of the
total cross section from 130 up to 250 MeV. The final set of parameters of
our dispersive relativistic optical model potential is summarized in Table
\ref{OMP_REL_PARAM}.

\subsection{Comparison with the experimental cross section in the energy
domain 0.1\mbox{$<$}E\mbox{$<$}250 MeV}

We now compare the experimental cross sections with those calculated from
our DOM potentials. The geometrical parameters of the model and the
strengths of the various components are specified in Tables \ref
{OMP_NOREL_PARAM} and \ref{OMP_REL_PARAM}. The dispersion relations fully
determine the real part of the dispersive contribution once the imaginary
part of the mean field is specified.

The $\sigma (\theta )$ relativistic DOM fits are compared to $^{27}Al(n,n)$
data in figure \ref{FIG_ANG_DISTR}. In general, the fits to $\sigma (\theta
) $ are of high quality. Very good agreement between experimental data and
calculations is observed in the energy region below 12 MeV, where CN
contribution is important. The highest deviation is observed for energies
25-26 MeV located near the diffraction maximum. In this energy region a
difficulty was encountered during the fit process evidenced by the fact that
a {\it common } set of surface absorptive potential parameters giving
acceptable fits to each type of data(differential and total cross section)
could not be found. The fits to $\sigma (\theta )$ indicate smaller values
of the imaginary surface potential depth $A_s$ parameter while fits to total
cross section point to a values larger by about 2 MeV. Experimental $\sigma
(\theta )$ data for energies higher than 26 MeV were not included in the
minimization procedure, but our relativistic DOM potential displays an
excellent agreement with these data.

The $\sigma _T$ relativistic DOM fit is compared with the total cross
section data and with calculations using phenomenological potentials in
figures \ref{FIG_SIGMA_HIGH_E} and \ref{FIG_SIGMA_LOW_E}. It should be
stressed that only the Finlay {\it et al}\cite{fifiab91,fiabfi93}
experimental total cross section data, shown as grey circles was used in the
DOM parameter search. In figure \ref{FIG_SIGMA_HIGH_E}, the total cross
section fit is in excellent agreement with the experimental data in the
whole energy range from 10 to 250 MeV. The only phenomenological potential
which gives a comparable agreement with experimental data up to 200 MeV is
the one by Koning and Delaroche\cite{kode01}, being slightly larger than
data in the region of the cross section maximum. Madland OMP overestimates
the experimental cross section by almost 20 \% above 100 MeV. The $\sigma _T$
relativistic DOM fit is compared to the high resolution total cross section
data measured by Ohkubo\cite{oh85} and Rohr {\it et al}\cite{roshna94} in
figure \ref{FIG_SIGMA_LOW_E}. The total cross section fit using relativistic
DOM potential is in good agreement with the averaged experimental data in
the whole energy range from 0.1 up to 10 MeV and practically equal to the
cross section derived from phenomenological OMP by Petler {\it et al}\cite
{peisfi85}. The total cross section calculated by the phenomenological
potential of Koning and Delaroche\cite{kode01} is smaller than the one
calculated by the relativistic DOM potential of the present work in the
whole energy range from 0.1 up to 10 MeV, but the shape remains quite
similar for all compared total cross section calculations. Calculation using
relativistic DOM potential, including reorientation effects by considering $%
Al$ as a deformed nucleus $(\beta =0.4)$ with ground state spin equal 2.5 is
shown as dashed line in the figure \ref{FIG_SIGMA_LOW_E}. This calculation
was carried out without readjusting any potential parameter to see the
effect of deformation on the total cross section. Maximum energy in this
calculation was equal to the energy of the first excited level to avoid
complexity linked to the coupled channel approach. We can see that
reorientation effects lead to the reduction of the calculated cross section
by 10\% from 0.1 up to 0.8 MeV. The small differences between the solid and
dashed curves are a measure of the error incurred by the neglect of
reorientation effects and nuclear deformation.

Figure \ref{FIG_POLAR_SIGMA} shows the comparison between the experimental
analyzing power and differential cross section at 7.5-7.6 MeV and the
predictions of our relativistic DOM. The agreement is good in view of the
fact that these data were not used in our fitting procedure. It can be seen
that CN contribution is still quite important at this energy. Polarization
measurements at 11 and 14 MeV are compared with DOM calculations in figure
\ref{FIG_POLAR}.

Average volume integral for the real part of the optical potential was
determined for the relativistic DOM potential as well as for the available
phenomenological potentials and is shown in figure \ref{FIG_REAL_VOLINT}. In
the same figure the ''Hartree-Fock'', volume and surface dispersive
contributions are shown. The biggest difference between our DOM potential
and the phenomenological ones are located below 50 MeV, where surface
dispersive contribution reach minimum and then changes the sign, becoming
positive. This pure dispersive effect can not be simulated by any variation
of the phenomenological OMP parameters. It is interesting that real volume
integral above 200 MeV is dominated by the dispersive volume contribution as
a result of the non-locality.

\vspace{0pt}Average volume integral for the imaginary part of the optical
potential was also calculated. In this case they are big differences between
phenomenological potentials and DOM results as can be seen in figure \ref
{FIG_IMAG_VOLINT}. Low energy behaviour is different as was the case for the
real volume integral, because the dominance of the dispersive contribution.
However high energy region is also quite different. DOM integral increases
with energy as a result of the non-locality contribution to the volume
absorptive potential. The only phenomenological potential showing similar
behaviour is the Madland OMP\cite{ma88}. His imaginary volume integral goes
parallel to the integral calculated using relativistic DOM potential (not
considering a discontinuity caused by two different functional forms
employed for reduced radius by Madland; one below 140 MeV, the second above
this value). There is a clear connection between this increase of the
imaginary volume integral and the saturation of the reaction cross section
at energies above 125 MeV as shown in figure \ref{FIG_SIGMA_REACT}. This
behaviour is consistent with the semiclassical estimation of the reaction
cross section. The relativistic DOM potential reaction cross section reaches
a near constant value of 0.3 barn. The asymptotical estimate of the reaction
cross section is $\pi R^2$ equivalent to the reduced radius of 1.03 fm. This
value compares well with the averaged reduced radius of 1.1-1.2 fm used for
the imaginary potential geometry of the DOM potential.\vspace{0pt} It is
interesting to point out that different reaction cross sections will have a
direct impact on cross sections available for any statistical model
calculations.

\section{Conclusions}

In this work we have presented a dispersive relativistic spherical optical
model analysis of neutron scattering up to 250 MeV for $^{27}$Al nucleus.
The excellent overall agreement obtained between predictions and
experimental data would not have been possible without including dispersive
terms in the calculations and non-locality effects in the volume absorptive
potential. New high precision scattering measurements for the aluminium
above 30 MeV are necessary to establish our analysis on firmer grounds and
confirm our present results.

\acknowledgements
This work was supported by Junta de Andaluc\'{\i }a and the spanish CICYT
under contract PB1998-1111 and by the European Union under contract
FIKW-CT-2000-00107. One of the authors (R.C.) acknowledges support from the
Ministerio de Educaci\'{o}n, Deportes y Cultura de Espa\~{n}a, Secretar\'{\i
}a de Estado de Educaci\'{o}n y Universidades.

\bibliographystyle{unsrt}
\bibliography{refer}

\newpage

\begin{center}
\begin{table}[!hbp]
\caption{Constant temperature level density parameters for residual nuclei
in $n+^{27}Al$ reaction}
\begin{tabular}{cccc}
Residual Nucleus & $E_{cut}$[MeV] & $T$[MeV] & $E_0$[MeV] \\ \hline
$^{27}$Al & 11.2 & 2.071 & -0.678 \\
$^{24}$Na & 5.2 & 1.875 & -2.046 \\
$^{27}$Mg & 6.0 & 2.113 & -1.2157
\end{tabular}
\label{LEV_DENS_PARAM}
\end{table}

\begin{table}[!hbp]
\caption{Optical model parameters for non-relativistic dispersive potential
for $n+^{27}Al$ reaction up to 150 MeV.}
\begin{tabular}{cc}
Parameter(Unit) & Value \\ \hline
$V_0$~(MeV) & 52.24 \\
$\alpha_{_{HF}}~(MeV^{-1})$ & 0.0071 \\
$A_v$~(MeV) & 12.5 \\
$B_v$~(MeV) & 58.8 \\
$r_v$~(fm) & 1.20 \\
$a_v$~(fm) & 0.65 \\
$A_s$~(MeV) & 12.6 \\
$B_s$~(MeV) & 3.25 \\
$C_s~(MeV^{-1})$ & 0.0395 \\
$r_s$~(fm) & 1.11 \\
$a_s$~(fm) & 0.64 \\
$E_F$~(MeV) & -10.392 \\
$E_p$~(MeV) & -5.66
\end{tabular}
\label{OMP_NOREL_PARAM}
\end{table}

\begin{table}[!hbp]
\caption{Optical model parameters for relativistic dispersive potential for
$n+^{27}Al$ reaction up to 250 MeV.}
\begin{tabular}{cc}
Parameter(Unit) & Value \\ \hline
$V_0$~(MeV) & 54 \\
$\alpha_{_{HF}}~(MeV^{-1})$ & 0.0087 \\
$A_v$~(MeV) & 7 \\
$B_v$~(MeV) & 65 \\
$r_v$~(fm) & 1.20 \\
$a_v$~(fm) & 0.63 \\
$A_s$~(MeV) & 12.5 \\
$B_s$~(MeV) & 5 \\
$C_s~(MeV^{-1})$ & 0.034 \\
$r_s$~(fm) & 1.11 \\
$a_s$~(fm) & 0.64 \\
$E_F$~(MeV) & -10.392 \\
$E_p$~(MeV) & -5.66
\end{tabular}
\label{OMP_REL_PARAM}
\end{table}
\end{center}

\newpage

\begin{figure}[tbp]
\caption{Dependence upon energy of the dispersive volume contribution of the
real central potential of the $n+^{27}Al$ mean field. The dotted curve
corresponds to equation (\ref{volumen}), in which it is assumed that the
imaginary part is symmetric about the Fermi energy. The thick solid curves
correspond to the asymmetric model, considering non-local behaviour of the
imaginary volume absorption above certain energy $E_a$ following equations (%
\ref{WVnonlocal1}) and (\ref{WVnonlocal2}). Thin dashed line corresponds to
the Fermi energy.}
\label{FIG_WDOM}
\end{figure}

\begin{figure}[tbp]
\caption{Cumulative number of levels as a function of the excitation energy
for the three residual nuclei considered in the CN cross section
calculations. The discrete level data are from the RIPL\protect\cite{RIPL98}
and are well represented by the "constant temperature" level density formula
of Ref. \protect\cite{gica65} using parameters from Table \ref
{LEV_DENS_PARAM}. Cut-off energy is indicated by the vertical dashed line.
Above the cut-off energy the "constant temperature" level density formula
was used}
\label{FIG_CUMUL_LEV}
\end{figure}

\begin{figure}[tbp]
\caption{Empirical real volume (solid circles) and imaginary volume
potential depth (empty circles) of the OMP for $n+^{27}Al$ as determined
from individual best $\chi ^2$ fit searches using $\sigma _{tot}$ data in
the interval $70<E<150$ MeV after the first iteration. (Upper panel) The
solid line for the Hartree-Fock potential is the functional representation
defined in equation (\ref{VHF}). The dashed line denotes the starting guess
values calculated using Lee {\it et al} OMP\protect\cite{lechfu99}. The
crosses represent the empirical values of the Hartree-Fock type potential
obtained after the dispersive contribution coming from the volume imaginary
part of the OMP was substracted from the real volume empirical values.
(Lower panel) The solid line for the absorptive potential is the functional
representation defined in equation (\ref{WV3}). The dashed line denotes the
starting guess values calculated using Lee {\it et al} OMP\protect\cite
{lechfu99}. }
\label{FIG_WVOL_FIT}
\end{figure}

\begin{figure}[tbp]
\caption{ Energy dependence of the $n+^{27}Al$ total cross section from 10
up to 150 MeV. The curve has been calculated using the non-relativistic
(solid line) DOM potential of the present work. Grey circles correspond to
Finlay {\it et al} \protect\cite{fifiab91,fiabfi93} experimental data used
in the fitting procedure. The diamonds, crosses and triangles are obtained
from the measurements by Taylor{\it et al}\protect\cite{tawo53}, Measday{\it %
et al}\protect\cite{mepa66} and Schneider{\it et al}\protect\cite{scco68}}
\label{FIG_SIGMA_HIGH_E_NONREL}
\end{figure}

\begin{figure}[tbp]
\caption{ Relativistic and nonlocality contribution to the total cross
section. The total cross section curves have been calculated using the
relativistic(solid line) and non-relativistic(dotted line) DOM potentials of
the present work. The dashed line denotes relativistic DOM potential results
without non-locality correction. }
\label{FIG_REL_NONLOCAL_CONTRIB}
\end{figure}

\begin{figure}[tbp]
\caption{Comparison between the neutron elastic differential cross section
experimental data and our DOM calculations(solid line). CN contributions has
been added to the direct reaction predictions for incident energies up to 12
MeV. The $\sigma (\theta )$ data were obtained from Towle and Gilboy%
\protect\cite{togi62} at 1,2,3 and 4 MeV; Tanaka {\it et al} \protect\cite
{tatsma69} at 4.8,6,7 and 8 MeV; Kinney and Perey\protect\cite{kipe70} at
5.4, 6.4,7.5 and 8.6 MeV; Dagge {\it et al}\protect\cite{dagrha89} at 7.62
MeV; Velkley {\it et al}\protect\cite{veglbr74} at 9 MeV; Boerker {\it et al}
\protect\cite{bobobr88} at 10.2 MeV; Whisnant {\it et al}\protect\cite
{whdago84} at 11,14 and 17 MeV and Petler {\it et al}\protect\cite{peisfi85}
at 18,20,22,25 and 26 MeV.It should be noted that data above 26 MeV was not
used in the fitting process. Neutron incident energy is quoted above each
calculated curve.}
\label{FIG_ANG_DISTR}
\end{figure}

\begin{figure}[tbp]
\caption{ Energy dependence of the $n+^{27}Al$ total cross section above 10
MeV. The curves have been calculated using the relativistic(solid line) DOM
potential of the present work. The dotted, dotted-dashed and dashed lines
have been obtained from the phenomenological OMP by Madland\protect\cite
{ma88}, Lee{\it et al}\protect\cite{lechfu99} and Koning and Delaroche%
\protect\cite{kode01} respectively in their range of validity. Grey circles
correspond to Finlay {\it et al}\protect\cite{fifiab91,fiabfi93}
experimental data used in the fitting procedure. The diamonds, crosses,
triangles, empty circles and solid squares have been obtained from the
measurements by Taylor{\it et al}\protect\cite{tawo53}, Measday{\it et al}%
\protect\cite{mepa66}, Schneider{\it et al}\protect\cite{scco68}, Franz{\it %
et al}\protect\cite{frgrle88} and Juren{\it et al}\protect\cite{demo51} }
\label{FIG_SIGMA_HIGH_E}
\end{figure}

\begin{figure}[tbp]
\caption{ Low energy dependence of the $n+^{27}Al$ total cross section from
0.1 up to 10 MeV. The curves have been calculated using the relativistic DOM
potential without(solid line) and with(dashed line) reorientation effects.
The circles, triangles up and triangles down have been obtained from the
phenomenological OMP by Harper{\it et al}\protect\cite{haal82}, Petler{\it %
et al}\protect\cite{peisfi85} and Koning and Delaroche{\it et al}%
\protect\cite{kode01}. The high resolution experimental data have been
obtained from the measurements by Ohkubo\protect\cite{oh85} and Rohr{\it et
al}\protect\cite{roshna94} }
\label{FIG_SIGMA_LOW_E}
\end{figure}

\begin{figure}[tbp]
\caption{The CN corrected $\sigma (\theta )$ and $A_y (\theta )$ data (solid
line) at $E_n=7.62$ MeV. Experimental data was taken from Dagge {\it et al}
\protect\cite{dagrha89} and Kinney and Perey\protect\cite{kipe70}. Dashed
line is denoted the uncorrected for CN contribution polarization and cross
section data.}
\label{FIG_POLAR_SIGMA}
\end{figure}

\begin{figure}[tbp]
\caption{The $A_y (\theta )$ data (solid line) at $E_n$=14 and 17 MeV.
Experimental data was taken from Martin and Walter\protect\cite{mawa86}.}
\label{FIG_POLAR}
\end{figure}

\begin{figure}[tbp]
\caption{Energy dependence of the volume integrals per nucleon of the
Hartree-Fock(dot-dashed line), volume(dotted line) and surface(dashed line)
dispersive components of the real part of the $n+^{27}Al$ mean field. The
thick solid curve represents the sum of all contributions. Nonlocality was
considered in the volume imaginary potential. The solid squares, solid
triangles and empty circles connected by lines have been obtained from the
phenomenological OMP by Madland\protect\cite{ma88}, Lee{\it et al}%
\protect\cite{lechfu99} and Koning and Delaroche\protect\cite{kode01}
respectively. }
\label{FIG_REAL_VOLINT}
\end{figure}

\begin{figure}[tbp]
\caption{Energy dependence of the volume integrals per nucleon of the volume
(dotted line) and surface-peaked(dashed line) components of the imaginary
part of the $n+^{27}Al$ mean field. The solid curve represents the sum of
all contributions. Nonlocality was considered in the volume imaginary
potential. The solid squares, solid triangles and empty circles connected by
lines have been obtained from the phenomenological OMP by Madland%
\protect\cite{ma88}, Lee{\it et al}\protect\cite{lechfu99} and Koning and
Delaroche\protect\cite{kode01} respectively. }
\label{FIG_IMAG_VOLINT}
\end{figure}

\begin{figure}[tbp]
\caption{ Energy dependence of the $n+^{27}Al$ reaction cross section from
0.1 up to 250 MeV. The thick solid curve has been calculated using the
relativistic DOM potential of the present work. The solid squares, solid
triangles and empty circles connected by lines have been obtained from the
phenomenological OMP by Madland\protect\cite{ma88}, Lee{\it et al}%
\protect\cite{lechfu99} and Koning and Delaroche\protect\cite{kode01}
respectively in their range of validity.}
\label{FIG_SIGMA_REACT}
\end{figure}

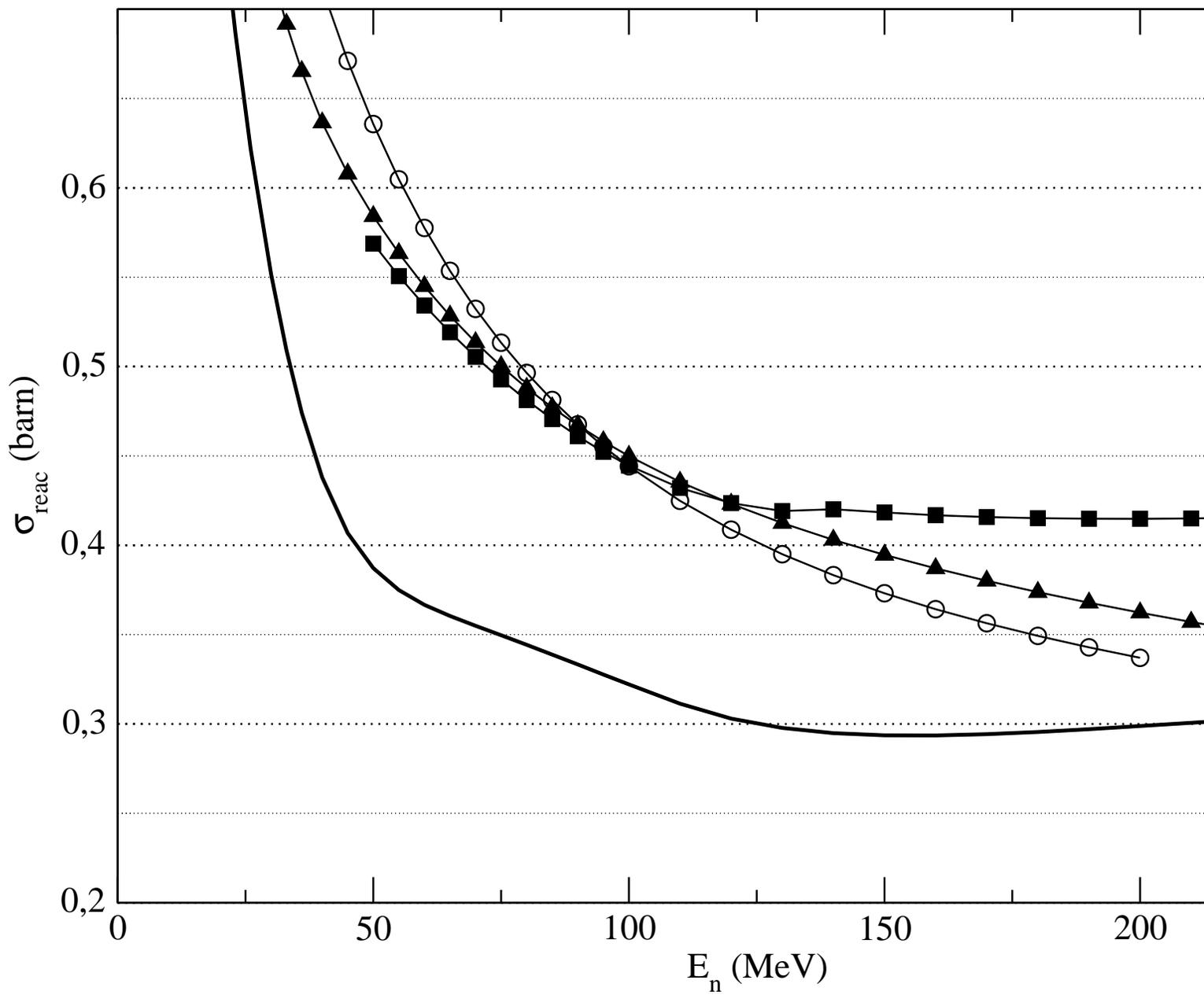
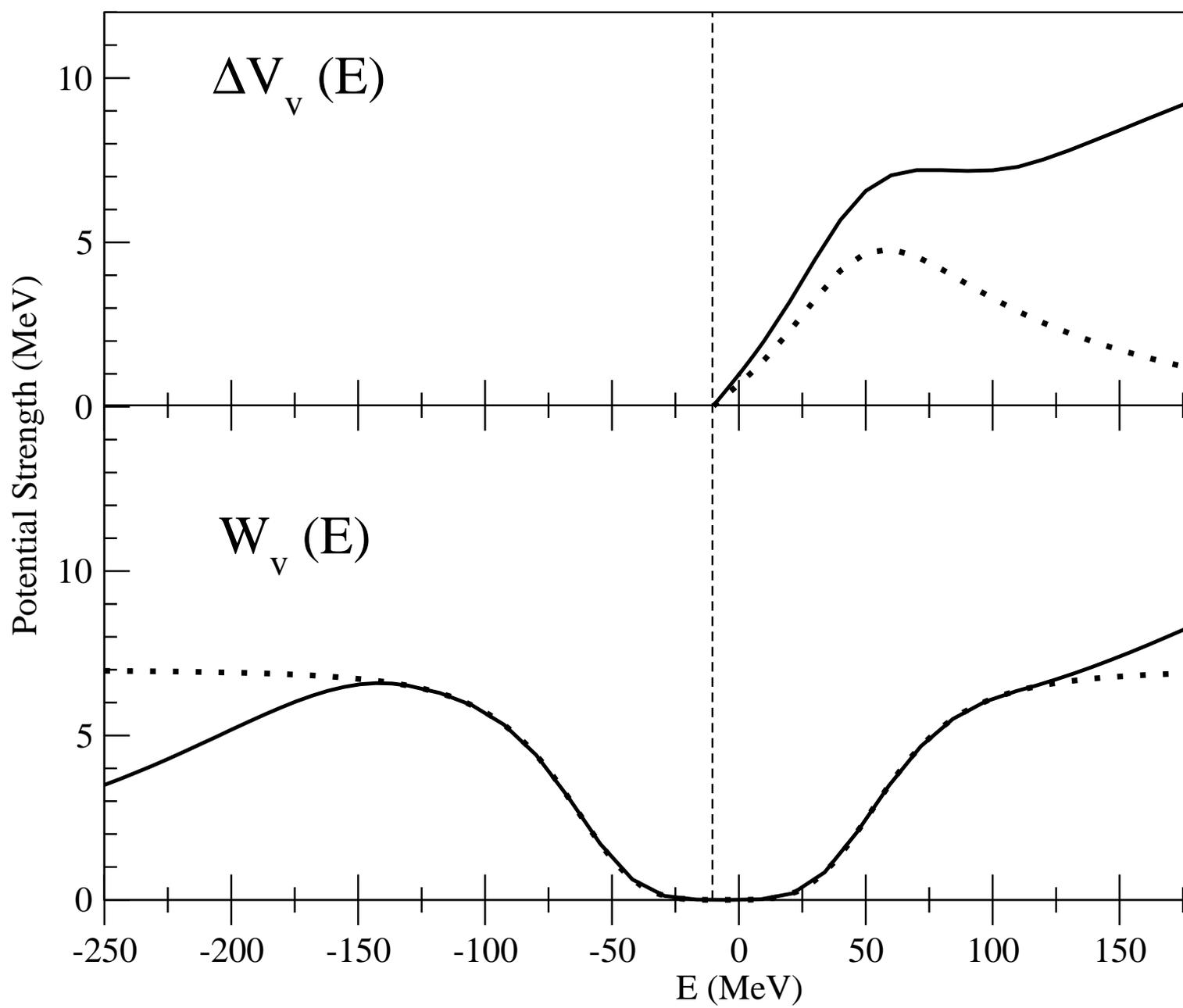
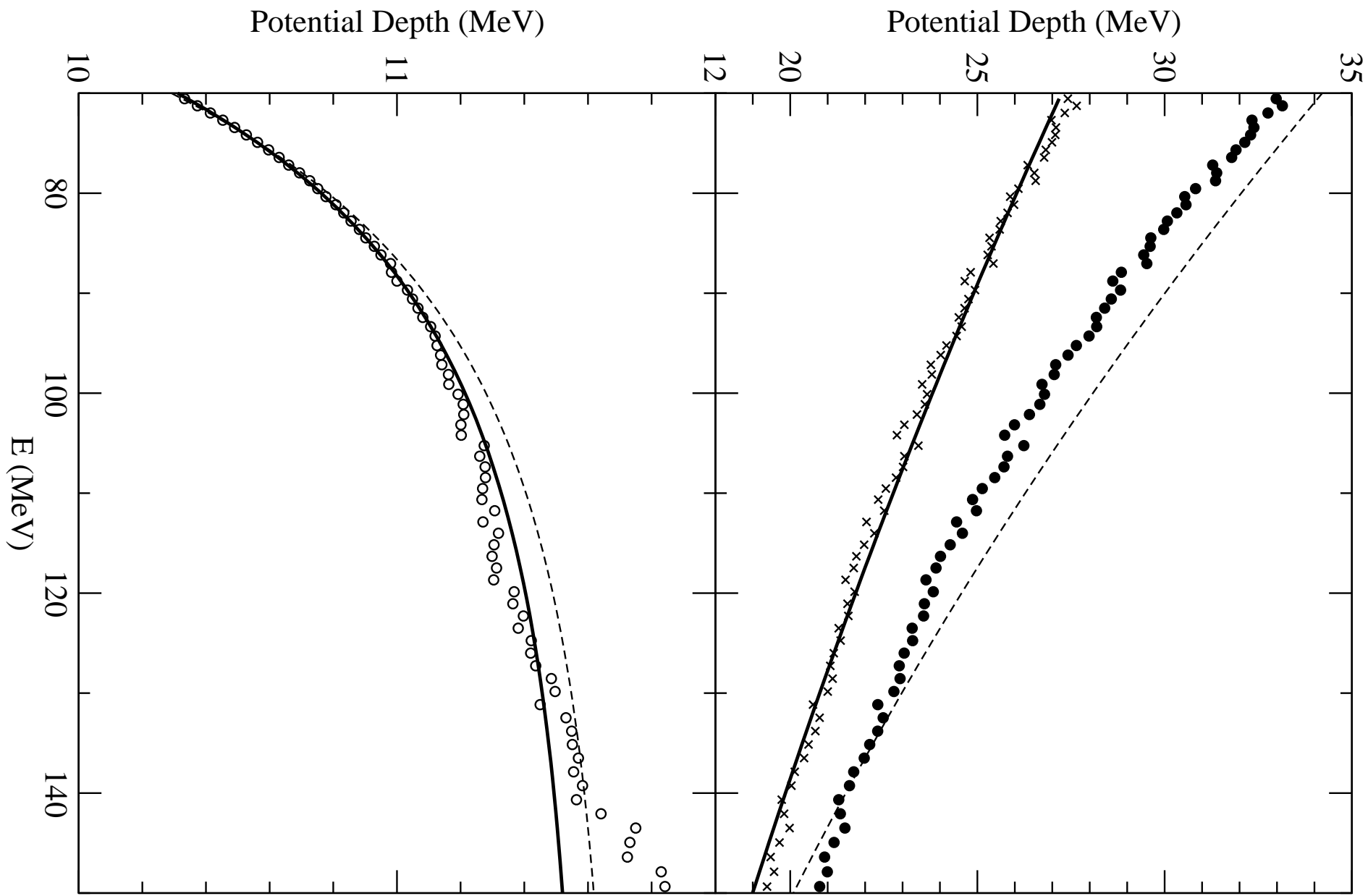
\end{document}